\documentclass[aps,prl,twocolumn,superscriptaddress]{revtex4-2}
\usepackage{graphicx}
\usepackage{mathrsfs}
\usepackage{bm}
\usepackage{amsmath}
\usepackage{color}
\usepackage{dcolumn}
\usepackage{epstopdf}
\usepackage{dsfont}
\usepackage{amssymb}
\usepackage{tabularx}
\usepackage{array,ulem}
\usepackage{float}
\usepackage{colordvi}
\usepackage{verbatim}
\usepackage{hyperref}

\hypersetup{
colorlinks=true,
linkcolor=blue,
filecolor=blue,
urlcolor=blue,
citecolor=blue,
}

\begin{document}
\title{Engineering Two-Dimensional Hybrid-Order Topological Insulators via Trilayer Coupling
%Interlayer-Engineered Two-dimensional Hybrid-Order Topological Insulator 
}

\author{Lizhou Liu}
\affiliation{International Center for Quantum Materials, School of Physics, Peking University, Beijing 100871, China}

\author{Cheng-Ming Miao}
\affiliation{International Center for Quantum Materials, School of Physics, Peking University, Beijing 100871, China}

\author{Qing-Feng Sun}
\email[Correspondence author:~~]{sunqf@pku.edu.cn}
\affiliation{International Center for Quantum Materials, School of Physics, Peking University, Beijing 100871, China}
\affiliation{Hefei National Laboratory, Hefei 230088, China}

\date{\today}

\begin{abstract}
We propose an interlayer-engineering scheme to realize a two-dimensional hybrid-order topological insulator, characterized by the coexistence of first-order and second-order topological phases, in a coupled trilayer Chern system. 
Starting from three quantum anomalous Hall layers with Chern numbers $\mathcal{C}_{1/2/3}=+1/-1/+1$ in the decoupled limit, interlayer tunneling hybridizes their edge states into a single chiral edge mode, while simultaneously opening a gap that supports corner states. 
Consequently, the system exhibits the coexistence of one-dimensional chiral edge states and zero-dimensional corner states within the same bulk gap, a hallmark of the hybrid-order topology.
Furthermore, we map out the topological phase diagram,  and show that the hybrid-order phase is robust against mass-type disorder. 
Our results identify interlayer hybridization as a minimal and broadly applicable strategy for engineering coexisting edge and corner states within a topological platform.
\end{abstract}

\maketitle

\textit{Introduction---}
Higher-order topological insulators extend the bulk-boundary correspondence of conventional first-order topological phases by supporting boundary modes of reduced dimensionality \cite{Hasan2010,Qi2011,Bansil2016,Kane2005,Bernevig2006,Schindler2018,Song2017,Langbehn2017,He2018,Liu2025QSH}.
In a $d$-dimensional bulk, an $n$th-order phase hosts protected $(d-n)$-dimensional boundary states.
Accordingly, two-dimensional (2D) first-order topological insulators can exhibit one-dimensional edge modes, whereas 2D second-order topological insulators can host zero-dimensional corner states \cite{Benalcazar2017,Benalcazar2017b, Ren2020,Liu2024,Liu2025a,Li2024,Yan2018, Mandal2025,Wang2025,Peng2025,Miao2024,Miao2023, Miao2025}.
In electronic systems, a central obstacle is that corner states are sharply localized and do not generically form conducting channels.
Therefore, diagnoses have mostly relied on band structures, local density of states, real-space imaging, or dynamical probes \cite{Qian2025,Liu2024Braiding,Liu2024HOTM, Miao2022}, and direct two-terminal signatures of localized corner states remain limited.
Although conductance features can be engineered by designing contact geometries to couple corner states to external leads \cite{Wang2021,Wang2024}, the resulting transport may involve hybridized scattering paths that become partially extended, which can obscure the geometric corner origin.

A natural strategy is to realize a 2D hybrid-order topological insulator, namely a bulk-gapped regime where a first-order edge channel and higher-order corner states coexist within the same bulk gap, so that edge-corner coupling is intrinsic.
Most discussions of hybrid-order topology focus on three-dimensional settings \cite{Bultinck2019,Kooi2020,Zhu2025,Yang2021,Hossain2024,Jia2025}, while a strict 2D counterpart remains much less explored.
Several recent works have considered 2D situations where edge and corner states appear simultaneously \cite{He2022, Yang2024,Lai2024,Yang2024a}.
However, these 2D settings are not strictly hybrid order in the above sense.
In particular, corner states can be effectively decoupled from conducting edge channels, for example by residing in a different bulk gap \cite{Yang2024,Lai2024} or by being separated into distinct spin sectors \cite{Yang2024a}.
Consequently, edge-corner scattering can be strongly reduced, and standard transport may not always provide a clear and universal signature of localized corner physics.
This motivates seeking a strict 2D hybrid-order phase, where an extended edge channel and corner states reside in the same bulk gap and can more directly intertwine.

\begin{figure}
  \centering
  \includegraphics[width=8.5 cm,angle=0]{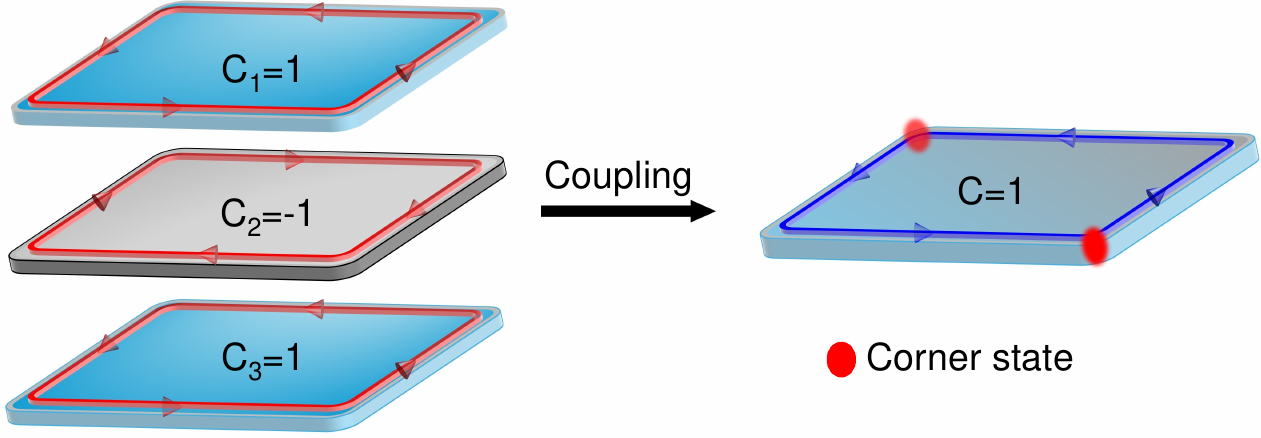}
  \caption{Schematic illustration of the interlayer-coupling mechanism. Left: three stacked Qi-Wu-Zhang layers with Chern numbers $\mathcal{C}_1=+1$, $\mathcal{C}_2=-1$, and $\mathcal{C}_3=+1$. Right: interlayer coupling hybridizes and gaps out two chiral edge channels, leaving a single blue chiral edge mode, red dots indicate the corner states.
}
  \label{fig1}
\end{figure}

In this Letter, we realize such a strict 2D hybrid-order topological insulator in a coupled trilayer Chern system and uncover its defining boundary phenomenology.
We start from three quantum anomalous Hall layers labeled by Chern numbers $\mathcal{C}_{1/2/3}=+1/-1/+1$ in the decoupled limit, for which each boundary hosts three chiral channels, two co-propagating and one counter-propagating,
as shown in Fig.~\ref{fig1}.
The interlayer tunneling hybridizes the boundary channels and gaps out two edge states, while a single chiral edge mode mandated by the total Chern number remains protected.
The resulting gapped boundary sector carries a nonzero mirror-graded winding number and binds corner states within the same energy window.
We map out the topological phase diagram, identify the phase boundaries, and show that the hybrid-order phase is robust against mass-type disorder.
Finally, using a two-terminal transport setup, we provide transport evidence that the corner states are coupled to the edge channel.

\textit{System Hamiltonian---}
We study a 2D trilayer Chern-insulator model formed by stacking three Qi-Wu-Zhang (QWZ) layers \cite{Qi2006} on a square lattice, as shown in Fig.~\ref{fig1}.
Each layer has two orbital degrees of freedom described by Pauli matrices $\sigma_{x,y,z}$ (with $\sigma_0$ the identity), and the lattice constant is set to unity.
In the decoupled limit, the top and bottom layers are identical, $H_1(\mathbf{k})=H_3(\mathbf{k})$, and we choose parameters such that their occupied bands carry $\mathcal{C}_1=\mathcal{C}_3=+1$.
The middle layer is designed to have the opposite Chern number $\mathcal{C}_2=-1$.

In momentum space, the coupled trilayer Bloch Hamiltonian is a $6\times 6$ matrix acting on the basis
$\Psi_{\mathbf{k}}=(c_{1\rho},c_{1\omega},c_{2\rho},c_{2\omega},c_{3\rho},c_{3\omega})^{T}$,
where $1,2,3$ label the layers and $\rho,\omega$ denote the two orbitals.
It takes the block form
\begin{equation}
H(\mathbf{k}) =
\begin{pmatrix}
H_{1}(\mathbf{k}) & t_{\perp} \,\sigma_0 & 0 \\
t_{\perp} \,\sigma_0 & H_{2}(\mathbf{k}) & t_{\perp} \,\sigma_0 \\
0 & t_{\perp} \,\sigma_0 & H_{3}(\mathbf{k})
\end{pmatrix},
\label{eq:H_trilayer}
\end{equation}
where, the intralayer blocks read
%\begin{equation}
\begin{eqnarray}
H_{1/3}(\mathbf{k})
 =  \sin k_x\,\sigma_x+\sin k_y\,\sigma_y
+ \bigl[m_1+\cos k_x+\cos k_y \bigr]\sigma_z , \nonumber
\label{eq:H13}\\
%\end{equation}
%\begin{equation}
H_2(\mathbf{k})
= \sin k_x\,\sigma_y+\sin k_y\,\sigma_x
+ \bigl[m_2+\cos k_x+\cos k_y \bigr]\sigma_z . \nonumber
\label{eq:H2}
%\end{equation}
\end{eqnarray}
And, $t_{\perp}$ denotes nearest-layer tunneling that preserves the orbital index.

\begin{figure}
  \centering
  \includegraphics[width=8.5 cm,angle=0]{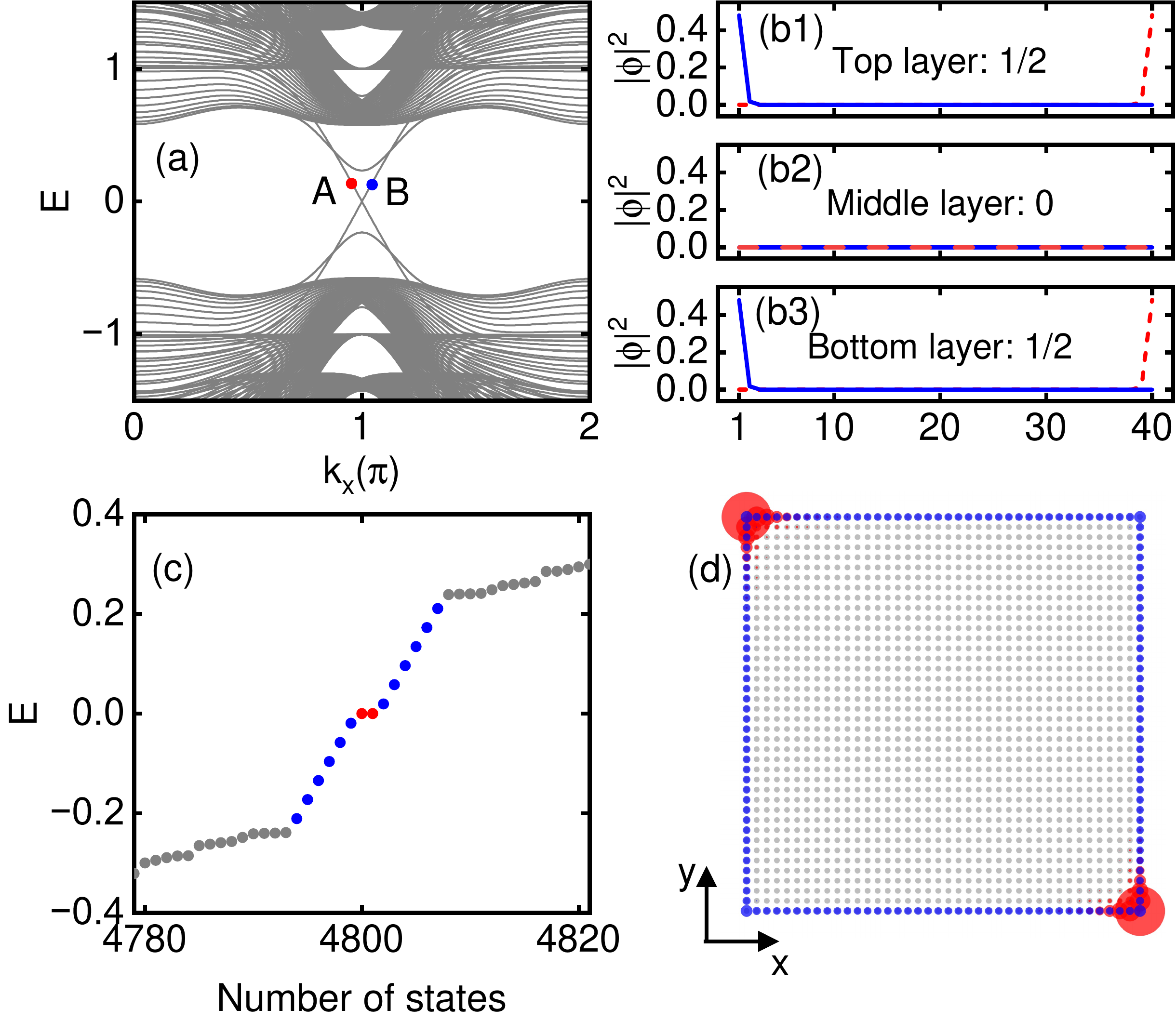}
  \caption{(a) Energy spectrum as a function of $k_x$ for a nanoribbon geometry. The labeled points $A$ and $B$ denote the two edge eigenstates.
(b1) -- (b3) Layer resolved probability distributions for the eigenstates at A (red dashed) and B (blue solid), shown for the top, middle, and bottom layers, respectively.
(c) Energy levels of a finite-size sample, where gapped edge states are shown in gray, gapless chiral edge states in blue, and the two near-zero corner states in red.
(d) Real-space probability density for the gapless edge states (blue) and the corner states (red) in (c). Parameters: $t_{\perp}=0.3$ and $m_1=m_2=1$.}
  \label{fig2}
\end{figure}

\textit{Hybrid-order boundary states---}
We first analyze a ribbon geometry and show the edge spectrum at a representative interlayer coupling $t_{\perp}=0.3$ in Fig.~\ref{fig2}(a).
Compared with the decoupled limit, interlayer tunneling hybridizes the boundary channels and gaps out two of them, while a single chiral edge mode remains gapless, as required by the total Chern number $\mathcal{C}=1$ of the coupled trilayer.
The labeled points $A$ and $B$ indicate representative eigenstates on this surviving gapless edge state.

Figures~\ref{fig2}(b1)–\ref{fig2}(b3) show the layer-resolved probability density $|\phi|^2$ of the edge eigenstates at the points $A$ and $B$ along the finite direction.
The surviving gapless chiral mode resides almost entirely on the top and bottom layers, each contributing approximately one half of the total edge weight, whereas the middle-layer component is negligible.
This layer selectivity reflects the interlayer hybridization, which gaps out the edge-channel combinations involving the middle layer and leaves the protected channel in the top-bottom antisymmetric sector.

We then turn to a rectangular nanoflake with open boundaries to diagnose the corner physics.
Figure~\ref{fig2}(c) shows the energy levels of this finite geometry, where the gapped edge states is shown in gray, the gapless chiral edge levels are highlighted in blue, and two nearly zero-energy states are highlighted in red.
As shown in Fig.~\ref{fig2}(d), the real-space probability density of a representative gapless edge state (blue) is extended along the boundary, whereas the two near-zero modes (red) are sharply localized at two opposite corners, establishing their corner-state character.
Moreover, the layer-resolved weights, $1/2$ on the middle layer and $1/4$ on each of the top and bottom layers, indicate that these corner states originate from the gapped edge sector.
Together, Fig.~\ref{fig2} demonstrates a 2D hybrid-order topological insulator in the coupled trilayer, where an extended chiral edge mode (blue) coexists with corner states (red) inside the same bulk gap.
Notably, the layer-resolved weights and real-space profiles show that the corner states spatially overlap with the boundary region, establishing a strict 2D hybrid-order phase.

In a finite sample, the spectrum is discrete and the chiral edge dispersion is sampled only at quantized boundary momenta. The resulting small near-zero level separation therefore reflects finite-size quantization rather than a true edge gap, as confirmed by the system-size scaling shown in Sec.~I and Fig.~S1 of the Supplemental Material (SM)~\cite{SM2025}. We further verify that the corner-state energies remain near zero and stay well separated from the other finite-size levels inside the bulk gap when the interlayer tunneling $t_{\perp}$ is varied within the hybrid-order regime, as shown in Sec.~II and Fig.~S2 of the SM~\cite{SM2025}. In addition, the zero-energy boundary-localized modes (i.e., corner states) persist under smooth boundary deformations that remove sharp corners, as shown in Sec.~III and Fig.~S3 of the SM~\cite{SM2025}, indicating that they are intrinsic topological modes rather than geometric artifacts.

To reveal the microscopic origin of this hybrid-order boundary phenomenology, we perform a reduction in the layer space. 
We work with the layer-resolved orbital spinors $\psi_\ell=(c_{\ell \rho},c_{\ell \omega})^{T}$ ($\ell=1,2,3$), so that the full basis reads $\Psi_{\mathbf{k}}=(\psi_1,\psi_2,\psi_3)$.
Using $H_{1}(\mathbf{k})=H_{3}(\mathbf{k})$ and the orbital-preserving interlayer tunneling $\propto\sigma_0$, we introduce the antisymmetric and symmetric layer combinations
$\psi_{a}=(\psi_{1}-\psi_{3})/\sqrt{2}$ and $\psi_{s}=(\psi_{1}+\psi_{3})/\sqrt{2}$.
In the basis $(\psi_a,\psi_s,\psi_2)$, the Hamiltonian becomes block diagonal,
$
(U^\dagger \otimes \sigma_0)\,H(\mathbf{k})\,(U\otimes \sigma_0)
=
H_{1}(\mathbf{k})
\oplus
\mathcal{H}_{s2}(\mathbf{k}),$
where the detailed derivation of the transformation is presented in Sec.~IV of the SM~\cite{SM2025} and
\begin{equation}
\mathcal{H}_{s2}(\mathbf{k})=
\begin{pmatrix}
H_{1}(\mathbf{k}) & \sqrt{2}\,t_{\perp}\sigma_0 \\
\sqrt{2}\,t_{\perp}\sigma_0 & H_{2}(\mathbf{k})
\end{pmatrix}.
\label{eq:Hs2_def_main}
\end{equation}
The antisymmetric block is exactly $H_1(\mathbf{k})$ and is completely decoupled from $t_\perp$, so it necessarily contributes the chiral edge channel as long as the bulk gap of $H_1$ remains open.
In contrast, $\mathcal{H}_{s2}$ represents the hybridized $(\psi_s,\psi_2)$ sector whose edge spectrum is gapped, yet it can be topologically nontrivial in a higher-order phase.

We find that, in the parameter regime of interest, the gapped sector $\mathcal{H}_{s2}$ carries a nontrivial second-order topology characterized by a nonzero mirror-graded winding number, which implies zero-energy corner states even though its edges are fully gapped \cite{Ren2020}.
This diagnosis follows from an additional diagonal-mirror symmetry that emerges for $m_1=m_2\equiv m$ along the line $k_x=-k_y$.
In the mirror eigenbasis, $\mathcal{H}_{s2}(k,-k)$ is block diagonalized into two decoupled $2\times2$ subsectors, each possessing a chiral symmetry and hence an integer winding number $\nu_{\pm}$.
In particular, we obtain $\nu_{+}=-\nu_{-}=1$, yielding $\nu=(\nu_{+}-\nu_{-})/2=1$.
By contrast, on the other diagonal $k_x=k_y$, the reduced Hamiltonian generically lacks a chiral symmetry, so the corresponding one-dimensional winding-number construction is not applicable. 
Consequently, this diagonal does not provide a symmetry-based topological diagnosis for boundary end states, and no symmetry-protected corner states arise along this direction. Further details of the symmetry analysis and the mirror-graded winding evaluation are provided in Sec.~V of the SM~\cite{SM2025}.
Consequently, in a rectangular nanoflake the zero-energy corner states appear only at the two corners connected by the $k_x=-k_y$ diagonal, consistent with Fig.~\ref{fig2}.
Moreover, since $\psi_s$ is an equal-weight superposition of the top and bottom layers, a corner mode originating from the $(\psi_s,\psi_2)$ sector naturally carries comparable weight on layers $(1,3)$ and on the middle layer, in agreement with Fig.~\ref{fig2}(d).
%Further details of the symmetry analysis and the mirror-graded winding evaluation are provided in Sec.~II of the SM~\cite{SM2025}.

\begin{figure}
  \centering
  \includegraphics[width=8.7 cm,angle=0]{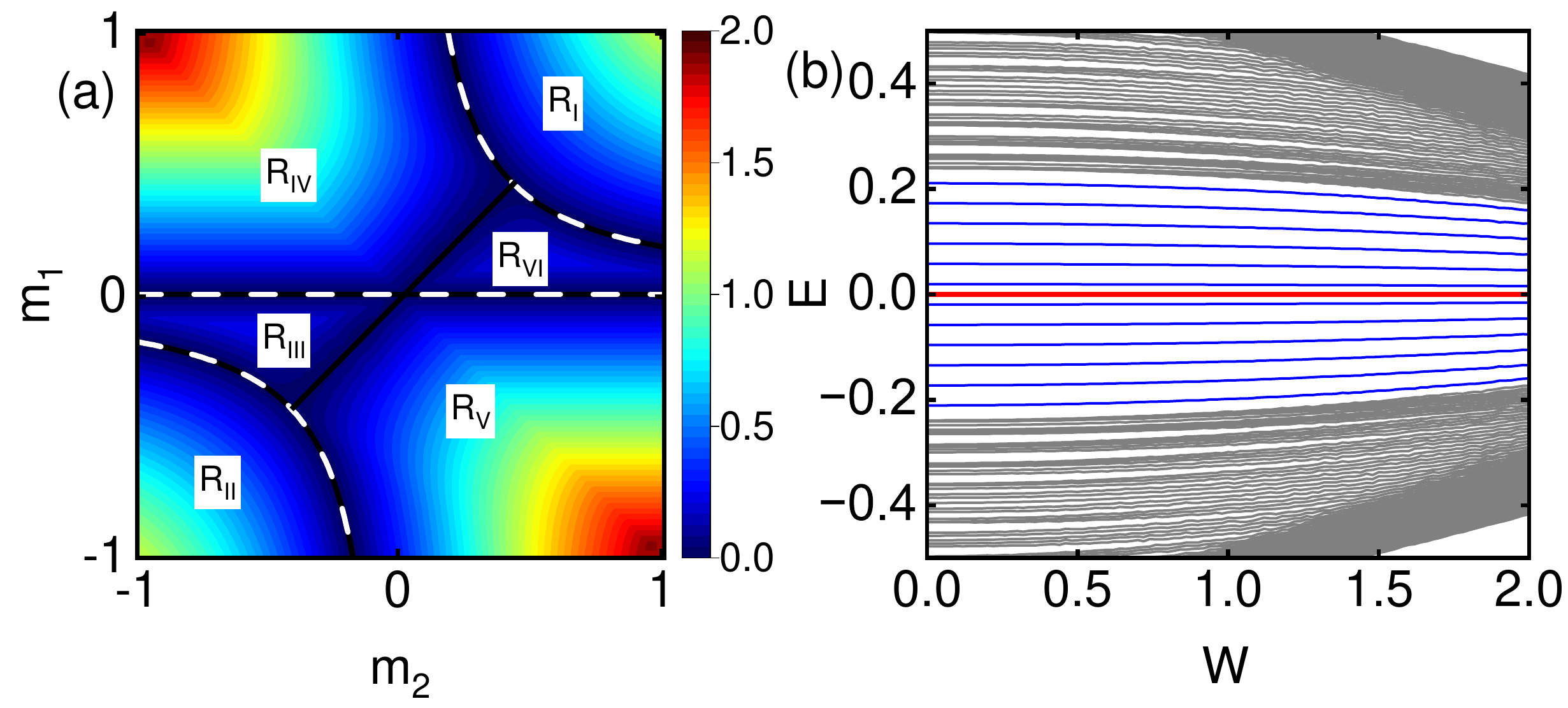}
  \caption{
(a) Phase diagram in the $(m_1,m_2)$ plane for $t_\perp=0.3$ with the bulk gap shown in color.
Solid black curves mark numerical gap closings, and white dashed curves show the analytical boundaries.
The labels $R_I-R_{VI}$ denote Regions~I--VI, with Regions~I and II corresponding to the hybrid-order phase. Outside Regions~I and II, the system is a $\mathcal{C}=1$ Chern insulator hosting three chiral boundary channels.
(b) Disorder-averaged finite-size spectrum versus mass-type disorder strength $W$, where blue lines highlight the gapless chiral edge states and the red line marks $E=0$ (corner states).
Parameters are the same as in Fig.~\ref{fig2}.
}
  \label{fig3}
\end{figure}

\textit{Phase diagram and robustness---}
To delineate the phase boundaries and identify the hybrid order regime, we track the minimum bulk gap in the $(m_1,m_2)$ plane. For fixed $t_{\perp}=0.3$, we compute the minimum bulk gap over the Brillouin zone and plot it in Fig.~\ref{fig3}(a). The resulting phase diagram shows distinct regions separated by gapless transition lines (black lines). Regions I and II in Fig.~\ref{fig3}(a) define the hybrid order regime, where a finite sample supports a single chiral edge mode together with a pair of corner states. 
Although the mirror-based winding-number diagnosis is strictly defined on the symmetric line $m_1=m_2$, the hybrid-order regime in Regions I and II of Fig.~\ref{fig3}(a) remains adiabatically connected to this limit because neither the bulk gap nor the edge gap closes when moving away from the symmetric line. This is explicitly verified by the nanoribbon and finite-size calculations shown in Sec.~VI and Fig.~S4 of the SM~\cite{SM2025}.

Outside Regions I and II, the system remains bulk gapped with $\mathcal{C}=1$, and a nanoribbon typically exhibits three edge modes. This is because the phase transitions in Fig.~\ref{fig3}(a) shift the momentum-space band inversion surface without changing the bulk Chern number~\cite{Wan2025}. 
Consequently, the edge modes on adjacent layers are displaced to different momenta, so the interlayer tunneling cannot efficiently hybridize and gap out the relevant edge sector, and robust corner states are absent. 
Representative nanoribbon spectra and layer-resolved edge profiles for Regions III to VI are provided in Sec.~VII and Fig.~S5 of the SM~\cite{SM2025}. They show that the edge states on adjacent layers are generally shifted in momentum and remain spatially distinct, which weakens the interlayer hybridization of the boundary sector and is consistent with the absence of robust corner states in these non-hybrid $\mathcal{C}=1$ regimes.

%Representative strip spectra for Regions III to VI are provided in Sec.~III and Fig.~S1 of the SM \cite{SM2025}, which directly show the momentum mismatch between the edge branches on adjacent layers and the resulting suppression of interlayer-hybridization-induced edge gapping in the non-hybrid $\mathcal{C}=1$ regime, leading to the emergence of three boundary channels in Regions III to VI.

We next rationalize the phase boundaries in Fig.~\ref{fig3}(a).
The layer-space reduction implies that a bulk gap closing can occur in either the antisymmetric block $H_{1}(\mathbf{k})$ or the gapped symmetric sector $\mathcal{H}_{s2}(\mathbf{k})$ defined in Eq.~(\ref{eq:Hs2_def_main}).
Within the parameter window $m_1\in(-1,1)$ displayed in Fig.~\ref{fig3}(a), band inversions are pinned to the high-symmetry points $\mathbf{k}_*\in\{X,Y\}$, and the gap closings at $X=(\pi,0)$ and $Y=(0,\pi)$.
Evaluating the gap-closing conditions $\det H_{1}(\mathbf{k}_*)=0$ and $\det \mathcal{H}_{s2}(\mathbf{k}_*)=0$ at $\mathbf{k}_*\in\{X,Y\}$ yields the analytic phase boundaries
$m_1=0$ and $m_1 m_2=2t_\perp^2$,
with the detailed derivation given in Sec.~VIII of the SM~\cite{SM2025}.
In Fig.~\ref{fig3}(a), the white dashed curves show these analytic boundaries and coincide with the numerically extracted gap-closing lines (solid black curves). This agreement validates the phase-boundary analysis and confirms that the transitions are controlled by symmetry-pinned band inversions at $X$ and $Y$.
We note that the additional solid black straight line arises from a gap closing at a generic momentum point away from $X$ and $Y$, for which an analytic expression is not available.
Equation $m_1m_2=2t_{\perp}^2$ also makes explicit that varying $t_{\perp}$ continuously deforms the phase boundaries, rather than introducing a finite-$t_{\perp}$ critical point. Consistently, $t_{\perp}=0$ corresponds to the decoupled limit, while the phase diagrams shown in Sec.~IX and Fig.~S6 of the SM~\cite{SM2025} demonstrate that the hybrid-order regime persists for finite $t_{\perp}$ and evolves continuously as $t_{\perp}$ varies.

To assess robustness against disorder, we consider onsite randomness in the orbital space,
$V_i=w_i\sigma_z$ with $w_i\in[-W/2,W/2]$ independently on each lattice site.
Figure~\ref{fig3}(b) shows the disorder-averaged finite-size spectrum versus $W$ for $m_1=m_2=1$ and $t_\perp=0.3$, which lie in the hybrid-order regime.
The two near-zero levels remain well separated from the other states over a broad range of $W$, indicating that the corner states are robust against this mass-type disorder.
This is consistent with the fact that $\sigma_z$ disorder mainly acts as a random mass and predominantly shifts local gaps, while it does not efficiently hybridize the corner levels with the propagating chiral edge continuum.
By contrast, for potential disorder $V_i=w_i\sigma_0$, the corner levels quickly broaden into the edge continuum, and the near-zero signature is already lost at weak disorder (see Sec.~X and Fig.~S7 of the SM~\cite{SM2025}).

\begin{figure}[t]
\centering
  \includegraphics[width=8.5 cm,angle=0]{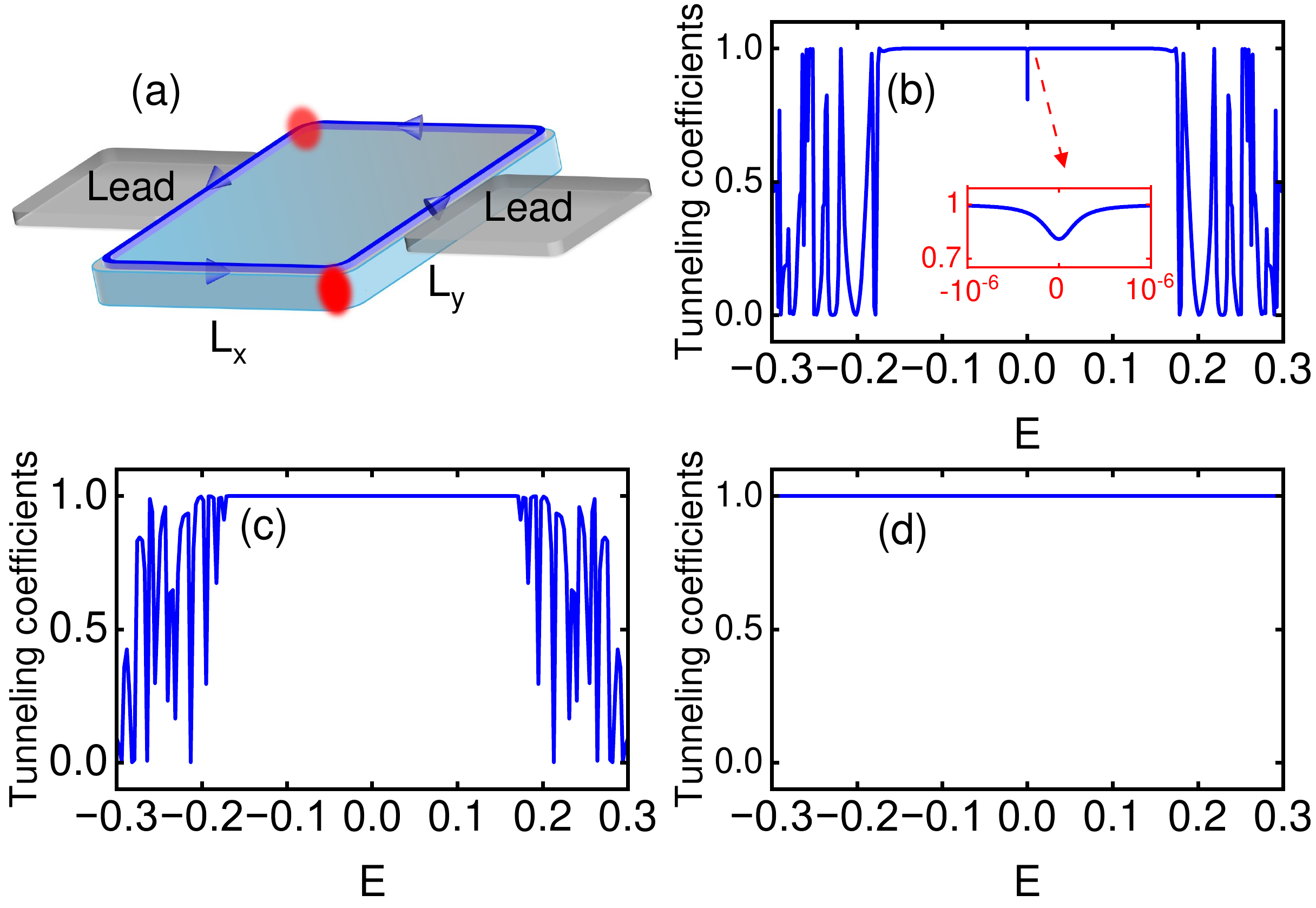}
\caption{
(a) Schematic of the device geometry.
A finite trilayer sample is connected to two semi-infinite leads (gray) that contact the top layer only.
The transport direction is along $x$ and the transverse direction is along $y$. 
The scattering region has length $L_x$ and width $L_y$, and each lead couples to an edge segment of width $W_{\mathrm{lead}}$ along the $y$ direction. 
Unless otherwise specified, $L_x=40$, $L_y=80$, and $W_{\mathrm{lead}}=40$. In panel (c), the wide-lead configuration corresponds to $W_{\mathrm{lead}}=L_y$.
(b) Energy-dependent transmission $T(E)$ for the setup in (a). The inset (red box) magnifies the dip region around $E=0$.
(c) Transmission $T(E)$ for a wide-lead configuration where the lead width matches the sample width.
(d) Transmission $T(E)$ in the decoupled-layer limit $t_{\perp}=0$.
Parameters are the same as in Fig.~\ref{fig2}.
}
\label{fig4}
\end{figure}

\textit{Transport evidence for corner-edge coupling---}
To seek further evidence that the corner sector is not fully decoupled from the conducting edge channel, we analyze a standard two-terminal transport geometry.
In the hybrid-order phase identified above, the corner levels lie within the same bulk gap as the surviving chiral edge channel and have finite weight near the boundary.
This motivates a Landauer-B\"uttiker calculation based on the Green-function formulation \cite{Landauer1957IBM,Buttiker1986PRL,FisherLee1981PRB}.
For a given incident energy $E$ (the Fermi energy), the coherent transmission from the left to the right lead is evaluated as
$
T(E)=\mathrm{Tr}\!\left[\Gamma_{\mathrm{L}}(E)\,G^{r}(E)\,\Gamma_{\mathrm{R}}(E)\,G^{a}(E)\right]
$
\cite{Meir1992PRL, Fisher1981}, where the trace is taken over the Hilbert space of the central scattering region, $G^{a}(E)=(G^{r}(E))^{\dagger}$, and
$
\Gamma_{\alpha}(E)=i\left[\Sigma_{\alpha}^{r}(E)-(\Sigma_{\alpha}^{r}(E))^{\dagger}\right]
$
with $\alpha=\mathrm{L},\mathrm{R}$.
Here $\Sigma_{\alpha}^{r}(E)$ is the retarded self-energy describing the coupling between the finite central region and the semi-infinite electrode $\alpha$.
The retarded Green's function of the central trilayer sample reads
\begin{equation}
G^{r}(E)=\left[(E+i\eta)I-H-\Sigma_{\mathrm{L}}^{r}(E)-\Sigma_{\mathrm{R}}^{r}(E)\right]^{-1},
\label{eq:Gr_transport}
\end{equation}
with $H$ the tight-binding Hamiltonian of the finite scattering region, $I$ the identity matrix in its Hilbert space, and $\eta>0$ an infinitesimal broadening ensuring the retarded boundary condition.
We take the left and right leads to be semi-infinite single-layer QWZ Chern insulators described by $H_1$, and the lead-sample coupling at the contacts is chosen to be identical to that inside the leads.
The leads are attached to the top layer only, so that $\Sigma_{\mathrm{L/R}}^{r}(E)$ has support exclusively in the top-layer subspace.
The electrode self-energies $\Sigma_{\mathrm{L/R}}^{r}(E)$ are computed from the surface Green's functions of the semi-infinite leads using standard recursive techniques \cite{Lee1981PRB}.

The transport setup is schematically shown in Fig.~\ref{fig4}(a).
The central trilayer scattering region has a size of $L_x=40$ (transport direction) and $L_y=80$ (transverse direction).
Two narrow leads of width $W_{\rm lead}=40$ are attached to the top layer.
As shown in Fig.~\ref{fig4}(b), in the coexistence regime the transmission forms an almost quantized plateau $T\simeq 1$ over the energy window of the surviving chiral edge channel, but it exhibits a clear suppression near zero energy.
The suppression originates from a corner-state-induced antiresonance: because the narrow leads are attached close to the corners, they couple strongly to the near-zero-energy corner levels. When the incident energy approaches a corner-state resonance, the localized corner mode acts as a side-coupled level rather than a conducting channel between the two leads, producing destructive interference in the transmission amplitude and thereby generating the dip in $T(E)$.
To further check that this suppression is tied to the corner sector, we perform two control calculations.
First, when the lead width is increased to match the sample width ($W_{\rm lead}=L_y$), the leads fully cover the transverse width of the scattering region, so no exposed contact corners remain at the lead-sample interfaces. Without these contact corners, corner-related localized states no longer form at the interfaces, and the corresponding transmission dip disappears, as shown in Fig.~\ref{fig4}(c).
Second, in the decoupled-layer limit $t_\perp=0$, the corner states are absent and the contacted top layer reduces to an ordinary Chern insulator, for which the transmission remains perfectly quantized without any suppression in the same energy window, as shown in Fig.~\ref{fig4}(d).
These comparisons indicate that the corner sector couples to the transport channel in this hybrid-order regime.

\textit{Discussion and conclusion---}
In summary, we have shown that a coupled trilayer Chern insulator realizes a 2D hybrid-order topological insulator, where a protected chiral edge channel coexists with corner states inside the same bulk gap.
We traced this hybrid-order boundary structure to an exact layer-space reduction, where an antisymmetric sector enforces the surviving chiral edge mode while the hybridized symmetric sector is characterized by a nonzero mirror-graded winding number that implies corner states.
In a standard two-terminal geometry, we find a transmission dip within the otherwise quantized plateau, suggesting that the corner sector is coupled to the propagating edge channel.
Taken together, these results support a strict 2D hybrid-order phase with edge--corner hybridization within a single bulk gap.

From a materials perspective, our proposal is broadly applicable and does not rely on fine-tuned microscopic details.
It requires only three layers realizing the parent Chern-number configuration $\mathcal{C}_{1/2/3}=+1/-1/+1$ in the decoupled limit, together with generic interlayer hopping.
Such ingredients can naturally arise in magnetic or altermagnetic van der Waals heterostructures and in engineered quantum anomalous Hall multilayers.
Recent first-principles studies have also reported second-order topological corner states in realistic bilayer Chern-insulator candidates, such as bilayer V$_2$WS$_4$ \cite{Liu2025AMHOTI}, supporting the plausibility of coupling-induced hybrid-order phases in trilayer systems.
More generally, higher-order phases in bilayer quantum spin Hall systems~\cite{Liu2025a} and bilayer topological superconductors~\cite{Miao2025a} can be viewed as extensions of coupled Chern-insulator constructions~\cite{Liu2024}.
The same stacking and hybridization strategy is therefore expected to carry over to trilayer quantum spin Hall and superconducting stacks, enabling helical edge modes with Kramers corner states or Majorana edge modes with Majorana corner states.

\textit{Acknowledgments---}
This work was financially supported by the National Key R and D Program of China (Grant No. 2024YFA1409002), the National Natural Science Foundation of China (Grants No.  12374034, No.  12547169, and No. 12447147), the Quantum Science and Technology-National Science and Technology Major Project (Grant No. 2021ZD0302403), and the China Postdoctoral
Science Foundation (Grant No. 2024M760070).
We also acknowledge the High-performance Computing Platform of Peking University for providing computational resources.


\begin{thebibliography}{100}
%\begin{comment}


\bibitem{Hasan2010} M. Z. Hasan and C. L. Kane, Colloquium: Topological insulators, Rev. Mod. Phys. \textbf{82}, 3045 (2010).
\bibitem{Qi2011} X.-L. Qi and S.-C. Zhang, Topological insulators and superconductors, Rev. Mod. Phys. \textbf{83}, 1057 (2011).
\bibitem{Bansil2016} A. Bansil, H. Lin, and T. Das, Colloquium: Topological band theory, Rev. Mod. Phys. \textbf{88}, 021004 (2016).
\bibitem{Kane2005} C. L. Kane and E. J. Mele, Quantum spin Hall effect in graphene, Phys. Rev. Lett. \textbf{95}, 226801 (2005).
\bibitem{Bernevig2006} B. A. Bernevig, T. L. Hughes, and S.-C. Zhang, Quantum spin Hall effect and topological phase transition in HgTe quantum wells, Science \textbf{314}, 1757 (2006).
\bibitem{He2018} K. He, Y. Wang, and Q.-K. Xue, Topological materials: Quantum anomalous Hall system, Annu. Rev. Condens. Matter Phys. \textbf{9}, 329 (2018).
\bibitem{Liu2025QSH} L. Liu, C.-M. Miao, Q.-F. Sun, and Y.-T. Zhang, Quantum spin Hall effect in bilayer honeycomb lattices with $C$-type antiferromagnetic order, {Phys. Rev. B \textbf{111}, 165421 (2025)}.




%% 高阶
\bibitem{Schindler2018} F. Schindler, A. M. Cook, M. G. Vergniory, Z. Wang, S. S. P. Parkin, B. A. Bernevig, and T. Neupert, Higher-order topological insulators, Sci. Adv. \textbf{4}, eaat0346 (2018).

\bibitem{Song2017} Z. Song, Z. Fang, and C. Fang, $(d\ensuremath{-}2)$-Dimensional Edge States of Rotation Symmetry Protected Topological States, Phys. Rev. Lett. \textbf{119}, 246402 (2017).

\bibitem{Langbehn2017} J. Langbehn, Y. Peng, L. Trifunovic, F. von Oppen, and P. W. Brouwer, Reflection-symmetric second-order topological insulators and superconductors, Phys. Rev. Lett. \textbf{119}, 246401 (2017).

\bibitem{addr1}
Y.-M. Li, Y.-J. Wu, X.-W. Luo, Y. Huang, and K. Chang,
Higher-order topological phases of magnons protected by magnetic crystalline symmetries, 
Phys. Rev. B \textbf{106}, 054403 (2022).

\bibitem{Benalcazar2017} W. A. Benalcazar, B. A. Bernevig, and T. L. Hughes, Quantized electric multipole insulators, Science \textbf{357}, 61 (2017).


\bibitem{Benalcazar2017b} W. A. Benalcazar, B. A. Bernevig, and T. L. Hughes, Electric multipole moments, topological multipole moment pumping, and chiral hinge states in crystalline insulators, Phys. Rev. B \textbf{96}, 245115 (2017).


\bibitem{Ren2020} Y. Ren, Z. Qiao, and Q. Niu, Engineering Corner States from Two-Dimensional Topological Insulators, {Phys. Rev. Lett. \textbf{124}, 166804 (2020)}.

\bibitem{Liu2024} L. Liu, J. An, Y. Ren, Y.-T. Zhang, Z. Qiao, and Q. Niu, Engineering second-order topological insulators via coupling two first-order topological insulators, {Phys. Rev. B \textbf{110}, 115427 (2024)}.


\bibitem{Liu2025a} L. Liu, J. An, Y. Ren, Y.-T. Zhang, Z. Qiao, and Q. Niu, Engineering corner states by coupling two-dimensional topological insulators, {Phys. Rev. B \textbf{111}, 045403 (2025)}.

\bibitem{Li2024} Y.-X. Li, Y. Liu, and C.-C. Liu, Creation and manipulation of higher-order topological states by altermagnets, {Phys. Rev. B \textbf{109}, L201109 (2024)}.

\bibitem{Yan2018} Z. Yan, F. Song, and Z. Wang, Majorana Corner Modes in a High-Temperature Platform, {Phys. Rev. Lett. \textbf{121}, 096803 (2018)}.

\bibitem{Mandal2025} S. Mandal, Z. Wang, R. Banerjee, H. T. Teo, M. Wei, P. Zhou, X. Xi, Z. Gao, G.-G. Liu, and B. Zhang, Photonic Bilayer Chern Insulator with Corner States,  {Phys. Rev. Lett. \textbf{135}, 016903 (2025)}.



\bibitem{Wang2025} L. Wang, A. H. Qureshi, Y. Sun, X. Xu, X. Yao, A.-L. He, Y. Zhou, and X. Zhang, Second-order topological insulator in bilayer borophene, {Phys. Rev. B \textbf{111}, 035113 (2025)}.


\bibitem{Peng2025} T. Peng, Y.-C. Xiong, X. Zhu, W. Cao, F. Lyu, Y. Hou, Z. Wang, and R. Xiong, Higher-order topological Anderson amorphous insulator, {Phys. Rev. B \textbf{111}, 075306 (2025)}.



\bibitem{Miao2024} C.-M. Miao, L. Liu, Y.-H. Wan, Q.-F. Sun, and Y.-T. Zhang, General principle behind magnetization-induced second-order topological corner states in the Kane-Mele model, {Phys. Rev. B \textbf{109}, 205417 (2024)}.

\bibitem{Miao2023} C.-M. Miao, Y.-H. Wan, Q.-F. Sun, and Y.-T. Zhang, Engineering topologically protected zero-dimensional interface end states in antiferromagnetic heterojunction graphene nanoflakes, {Phys. Rev. B \textbf{108}, 075401 (2023)}.

\bibitem{Miao2025} C.-M. Miao, Y.-H. Wan, Y.-T. Zhang, and Q.-F. Sun, Tunable second-order topological corner states induced by interlayer coupling in twisted bilayer Chern insulators, Phys. Rev. B \textbf{111}, 165418 (2025).

\bibitem{Qian2025} Y. Qian et al., Programmable higher-order nonequilibrium topological phases on a superconducting quantum processor, Science \textbf{390}, 930 (2025).


\bibitem{Liu2024Braiding} L. Liu, C.-M. Miao, H. Tang, Y.-T. Zhang, and Z. Qiao, Magnetically controlled topological braiding with Majorana corner states in second-order topological superconductors, {Phys. Rev. B \textbf{109}, 115413 (2024)}.


\bibitem{Liu2024HOTM} L. Liu, C.-M. Miao, Q.-F. Sun, and Y.-T. Zhang, Two-dimensional higher-order topological metals, {Phys. Rev. B \textbf{110}, 205415 (2024)}.

\bibitem{Miao2022} C.-M. Miao, Q.-F. Sun, and Y.-T. Zhang, Second-order topological corner states in zigzag graphene nanoflake with different types of edge magnetic configurations, {Phys. Rev. B \textbf{106}, 165422 (2022)}.
% 角态 dimer（Sci. China 2021）
\bibitem{Wang2021} K.-T. Wang, Y. Ren, F. Xu, Y. Wei, and J. Wang, Transport induced dimer state from topological corner states, Sci. China Phys. Mech. Astron. \textbf{64}, 257811 (2021).


% Spin separation and filtering via corner states（PRB 110, 125433）
\bibitem{Wang2024} K.-T. Wang, H. Wang, S. Liu, M. Wei, and F. Xu, Spin separation and filtering assisted by topological corner states in the Kekul\'e lattice, Phys. Rev. B \textbf{110}, 125433 (2024).

% 三维混合阶
\bibitem{Bultinck2019} N. Bultinck, B. A. Bernevig, and M. P. Zaletel, Three-dimensional superconductors with hybrid higher-order topology, Phys. Rev. B \textbf{99}, 125149 (2019).

\bibitem{Kooi2020} S. H. Kooi, G. van Miert, and C. Ortix, Hybrid-order topology of weak topological insulators, Phys. Rev. B \textbf{102}, 041122(R) (2020).

\bibitem{Zhu2025} Y.-Q. Zhu, Z. Zheng, G. Palumbo, and Z. D. Wang, Topological insulators with hybrid-order boundary states, Phys. Rev. B \textbf{111}, 195107 (2025).

\bibitem{Yang2021} Y. Yang, J. Lu, M. Yan, X. Huang, W. Deng, and Z. Liu, Hybrid-order topological insulators in a phononic crystal, Phys. Rev. Lett. \textbf{126}, 156801 (2021).

\bibitem{Hossain2024} M. S. Hossain et al., A hybrid topological quantum state in an elemental solid, Nature \textbf{628}, 527 (2024).

\bibitem{Jia2025}  W. Jia, Y. Tian, H. Yang, X. Kong, Z.-H. Huang, W.-J. Gong, and J.-H. An, Unconventional hybrid-order topological insulators, Phys. Rev. B \textbf{112}, L241103 (2025).


\bibitem{He2022} A.-L. He, W.-W. Luo, Y. Zhou, Y.-F. Wang, and H. Yao, Topological states in a dimerized system with staggered magnetic fluxes, Phys. Rev. B \textbf{105}, 235139 (2022).


% 二维  不同gap
\bibitem{Yang2024} N.-J. Yang, Z. Huang, and J.-M. Zhang, Hybrid-order topological phase and transition in 1$H$ transition metal compounds, Appl. Phys. Lett. \textbf{125}, 263102 (2024).

\bibitem{Lai2024} P. Lai, J. Wu, Z. Pu, J. Lu, H. Liu, W. Deng, H. Cheng, S. Chen, and Z. Liu, Real-projective-plane hybrid-order topological insulator realized in phononic crystals, Phys. Rev. Applied \textbf{21}, 044002 (2024).
% 二维  不同spin
\bibitem{Yang2024a} N.-J. Yang, J.-M. Zhang, X.-P. Li, Z. Zhang, Z.-M. Yu, Z. Huang, and Y. Yao, Sliding Ferroelectrics Induced Hybrid-Order Topological Phase Transitions, Phys. Rev. Lett. \textbf{134}, 256602 (2025).



\bibitem{Qi2006} X.-L. Qi, Y.-S. Wu, and S.-C. Zhang, Topological quantization of the spin Hall effect in two-dimensional paramagnetic semiconductors, Phys. Rev. B \textbf{74}, 085308 (2006).


\bibitem{SM2025} See Supplemental Material at [URL will be inserted by publisher] for the layer-space reduction, diagonal-mirror and winding analysis of $\mathcal{H}_{s2}$, nanoribbon spectra in Regions III--VI, analytic gap-closing phase boundaries, and onsite-disorder robustness.



\bibitem{Wan2025} Y.-H. Wan, P.-Y. Liu, and Q.-F. Sun, Classification of Chern numbers based on high-symmetry points, Phys. Rev. B \textbf{111}, L161410 (2025).


\bibitem{Landauer1957IBM} R. Landauer, Spatial variation of currents and fields due to localized scatterers in metallic conduction, IBM J. Res. Dev. \textbf{1}, 223 (1957).

\bibitem{Buttiker1986PRL} M. B\"uttiker, Four-terminal phase-coherent conductance, Phys. Rev. Lett. \textbf{57}, 1761 (1986).
\bibitem{FisherLee1981PRB} D. S. Fisher and P. A. Lee, Relation between conductivity and transmission matrix, Phys. Rev. B \textbf{23}, 6851(R) (1981).


\bibitem{Meir1992PRL}
Y.~Meir and N.~S.~Wingreen,
Landauer formula for the current through an interacting electron region,
Phys. Rev. Lett. \textbf{68}, 2512 (1992).


\bibitem{Fisher1981}  D. S. Fisher and P. A. Lee, Relation between conductivity and transmission matrix, Phys. Rev. B \textbf{23}, 6851(R) (1981).


\bibitem{Lee1981PRB}
D.~H.~Lee and J.~D.~Joannopoulos,
Simple scheme for surface-band calculations. II. The Green's function,
Phys. Rev. B \textbf{23}, 4997 (1981).


\bibitem{Liu2025AMHOTI} K. Liu and M. Zhao, Altermagnetism and higher-order topological states in bilayer Chern insulators, Phys. Rev. B \textbf{112}, L241405 (2025).

\bibitem{Miao2025a} C.-M. Miao, Y.-H. Wan, Y.-T. Zhang, and Q.-F. Sun, Tunable Majorana corner states driven by superconducting phase bias in a vertical Josephson junction, Phys. Rev. B \textbf{112}, 094507 (2025).


\end{thebibliography}
\end{document}